%% file: s1.tex
\documentclass{article}
\usepackage{spconf,amsmath,graphicx}
\input{preamble}

\title{Signal processing with a distribution of graph operators}
\name{ Feng Ji \qquad Wee Peng Tay}
\address{School of Electrical and Electronic Engineering, Nanyang Technological University, Singapore}
\begin{document}
\ninept
\maketitle
\begin{abstract}
In this paper, we develop a signal processing framework of a network without explicit knowledge of the network topology. Instead, we make use of knowledge on the distribution of operators on the network. This makes the framework flexible and useful when accurate knowledge of graph topology is unavailable. Moreover, the usual graph signal processing is a special case of our framework by using the delta distribution. The main elements of the theory include Fourier transform, theory of filtering and sampling.
\end{abstract}

\section{Introduction}

Since its emergence, the theory and applications of graph signal processing (GSP) have rapidly developed \cite{Shu13, San13, San14, Gad14, Don16, Egi17, Sha17, Gra18, Ort18, Girault2018, Ji19}. The theory bases on the choice of a symmetric \emph{graph shift operator} $S$, which is a preferred linear transformation on the vector space of graph signals. Once such an operator is given, there is a systemic way to develop a framework for graph signal processing. To highlight a few important elements of GSP, change of basis w.r.t.\ an eigenbasis of $S$ defines the \emph{graph Fourier transform}. The coefficients of a graph signal in the new basis are the components in the \emph{frequency domain}. \emph{Convolution} is just transformation by a diagonal matrix in the frequency domain. \emph{Sampling} refers to finding pseudo-inverse of a special family of convolutions, namely, the band-pass filters. 

Common choices of the graph shift operator include the graph adjacency matrix, the graph Laplacian and their normalized versions. However, knowing such a graph shift operator is equivalent to having complete knowledge of graph topology. Such knowledge may not be readily available in many real scenarios, and learning graph topology is itself a challenging task. 

For example, in a sensor network, physical distance between sensors are usually known. In order to obtain a sensor network with reasonable sparsity, only those connections between sensors close enough are included in the network, using approaches such as the $k$-nearest neighbors ($k$-NN) \cite{Alt92, San13}. In another common scenario \cite{Egi17, Don19, Mat19}, there are features associated with each member of a network. The graph is constructed based on similarities of such features, which makes construction accuracy depends largely on the features. In the case of information transmission over a network, the effective network topology depends not only on physical connections but also on transmission rate between members of the network. Certain topology inference methods are sensitive to accuracy of knowledge of such rate \cite{JiT19}. In the recent work \cite{JGT20}, GSP is extended to signal processing over simplicial complexes as a high dimensional generalization, which also involves an inference of the simplicial structure. 

In this paper, we propose a signal processing framework that forgoes the step of choosing a graph shift operator. Instead, we consider a distribution of operators on signals living on a finite discrete set $V$. We introduce the basic setup and define Fourier transform in \cref{sec:dis}. In \cref{sec:conv} - \cref{sec:ba}, we discuss various families of filters. We present sampling theory in \cref{sec:ban} in conjunction with band-pass filters. We show simulation results in \cref{sec:sim} and conclude in \cref{sec:con}.  

\section{Distribution of graph operators} \label{sec:dis}

Let $V$ be a finite discrete set of points, understood as of members of a network. A signal on $V$ is a function $f: V \to \mathbb{R}^n, n=|V|$, where each signal associates a number to a vertex $v\in V$. Denote the space of such signals by $L^2(V)$. Suppose $X$ is a set of operators on $V$, i.e., each $x\in X$ is a positive semi-definite symmetric matrices of size $|V|$. Assume that $(X, F, \mu_X)$, or abbreviated by $(X,\mu_X)$ if the $\sigma$-algebra $F$ is clear from the context, is a probability space. In many cases, we assume that $\mu_X$ has a density function $p$, i.e., $\mu_X=pdx$ where $dx$ is a base measure. We call $X$ the \emph{base space}. Suppose we have a product $X\times Y$. Then $x\times Y, x\in X$ is called the \emph{fiber} at $x$. 

To understand such a setup, we consider the example that there is a distribution of graphs on $V$. Each $G$ from the distribution gives a Laplacian $L_G$, and the collection of such $L_G$ yields the set $X$. An even more specific case is when there is an unweighted graph $G$ on $V$ that describes the connections among $V$, and the distribution comes from a distribution of edge weights.

\begin{Definition} \label{defn:dbn}
	Denote by $[n]$ the discrete set $\{1,\ldots, n\}$. For each $x \in X$, let $\lambda_x(i)$ be the $i$-th eigenvalue of the operator of $x$ (ordered increasingly) and $v_x(i)$ be the associated eigenvector. Define the \emph{Fourier transform} w.r.t.\ $(X,\mu_X)$ such that $\mu_X=pdx$ as
	\begin{align*} 
	\phi_{X}: L^2(V) \to L^2(X\times [n])
	\end{align*}
	by $f \mapsto \hat{f}$ where $\hat{f}(x,i) = p(x)^{1/2}\langle f, v_x(i)\rangle$.
\end{Definition} 

$L^2(V)$ is nothing but an $n$-dimensional vector space. On the other hand, $L^2(X\times [n])$ can be infinite dimensional in general. Therefore, $\phi_X$ cannot be invertible. However, it has a left inverse:
\begin{align*}
\psi_X: L^2(X\times [n]) \to L^2(V), \\
g \mapsto \int_x \sum_{i=1}^n g(x,i)p(x)^{1/2}v_x(i)dx.
\end{align*}
We have the following basic properties. 

\begin{Lemma}
	\begin{enumerate}[(a)]
		\item $\phi_X$ and $\psi_X$ are both well-defined; and $\psi_X \circ \phi_X$ is the identity map on $L^2(V)$.
		\item (Parseval's identity) $\norm{f} = \norm{\phi_X(f)}$ for $f\in L^2(V)$. 
	\end{enumerate}
\end{Lemma}

The coefficient $p(x)^{1/2}$ in defining the Fourier transform is essential if the space $X$ does not have finite base measure.

\begin{Example}
	\begin{enumerate}[(a)]
	\item Suppose $\mu_X = \delta_x$ is the delta distribution on $x\in X$. Then $\phi_X$ is just the Fourier transform in GSP theory w.r.t.\ the operator $x$.
	 
	\item Consider that there is a graph $G$ that describes connections among $V$. Suppose we have two Laplacians $L_1$ and $L_2$ representing two subgraph types on $G$, and $X$ is parametrized by the unit interval $T = [0,1]$. For each $t \in [0,1]$, we form $L_t = tL_1+(1-t)L_2$. Here, $F$ is the Borel sigma algebra and $\mu_X$ is a probability measure on $[0,1]$. If $\mu_X$ is absolute continuous w.r.t.\ the Lebesgue measure, then there is a density $p(x)$. If we have prior knowledge on $p(x)$, then we may use Fourier transform to analyze signals on $G$, without explicit knowing a particular instance of a shift operator.
	
	We show in \figref{fig:het1} Fourier transform of signals with $L_1$ and $L_2$ being a random split of the Laplacian $L$ of a graph $G$. The signal feature and $\mu_X$ are captured in the spectral plots.
	
	\begin{figure}[!htb] 
		\centering
		\includegraphics[width=0.33\columnwidth,trim=0 6cm 0 6cm, clip]{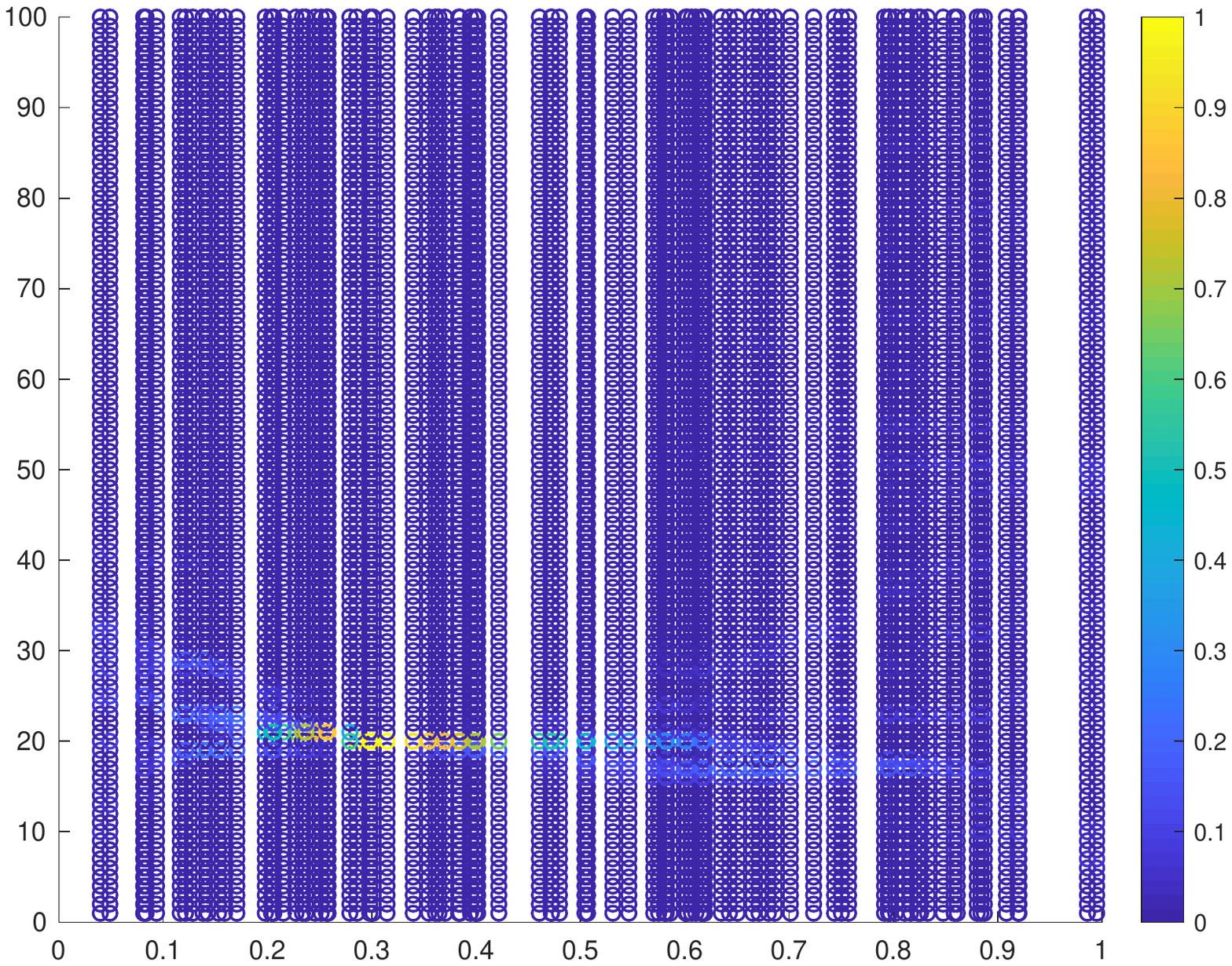}
		\includegraphics[width=0.33\columnwidth,trim=0 6cm 0 6cm, clip]{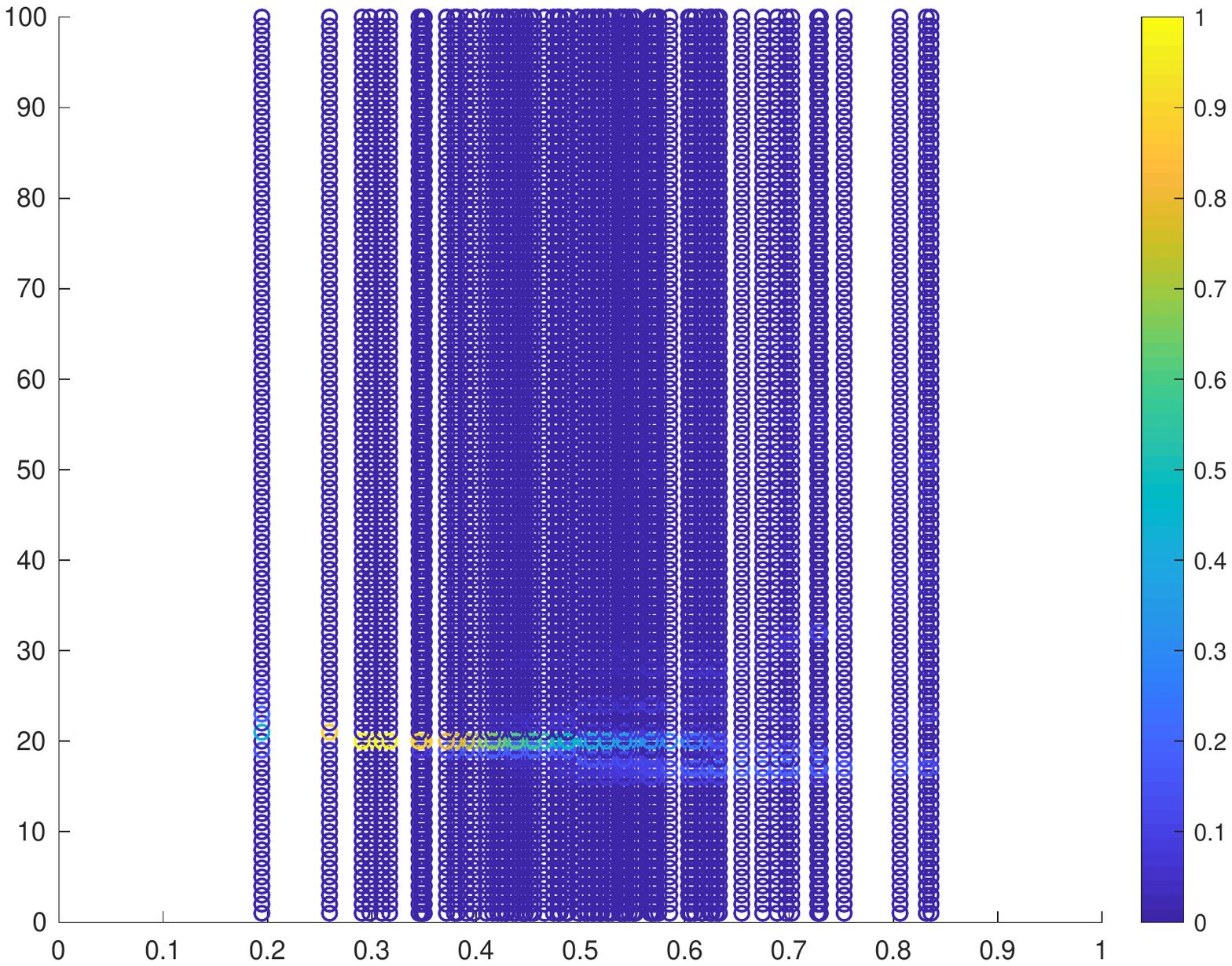}
		\includegraphics[width=0.33\columnwidth,trim=0 6cm 0 6cm, clip]{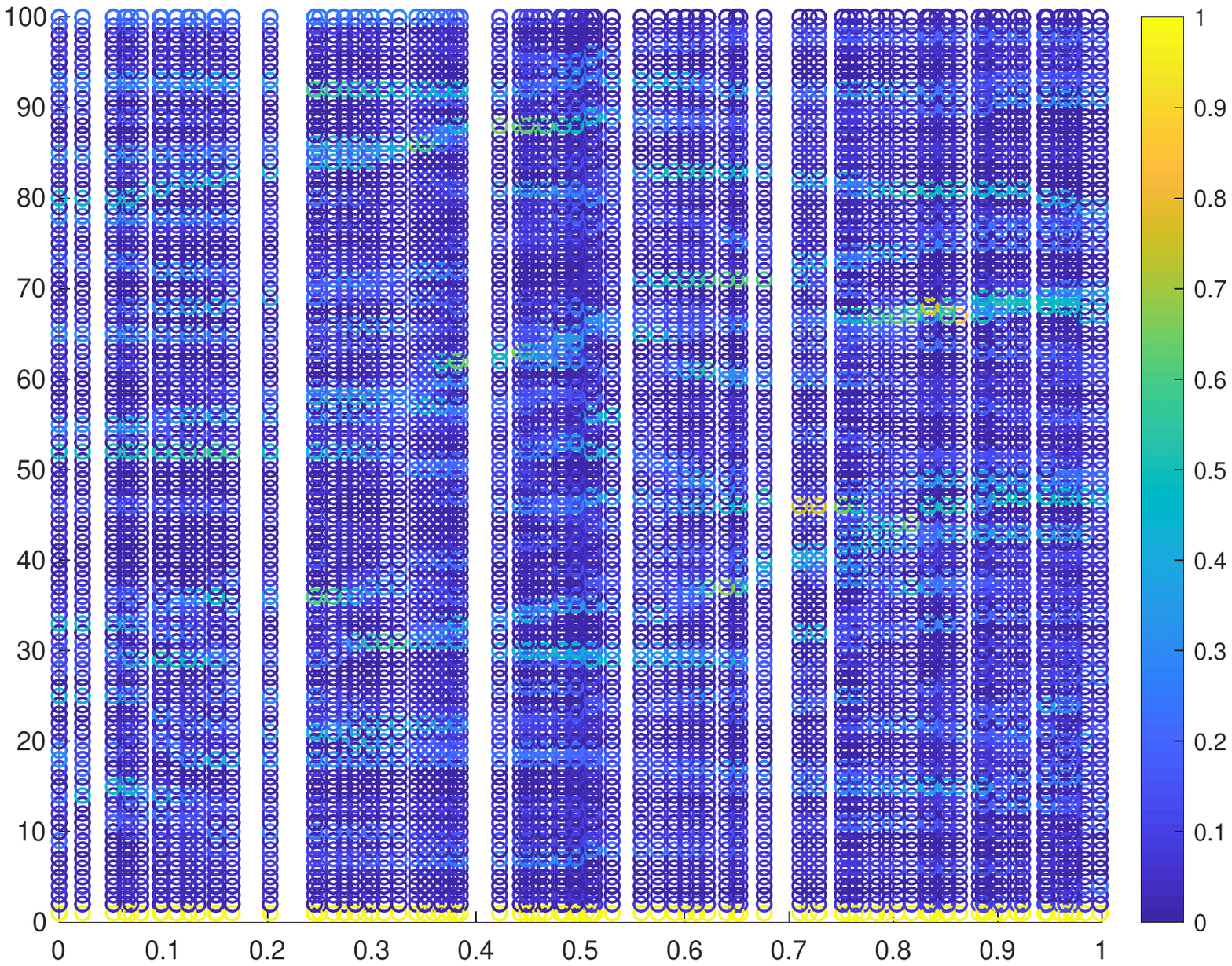}
		\includegraphics[width=0.33\columnwidth,trim=0 6cm 0 6cm, clip]{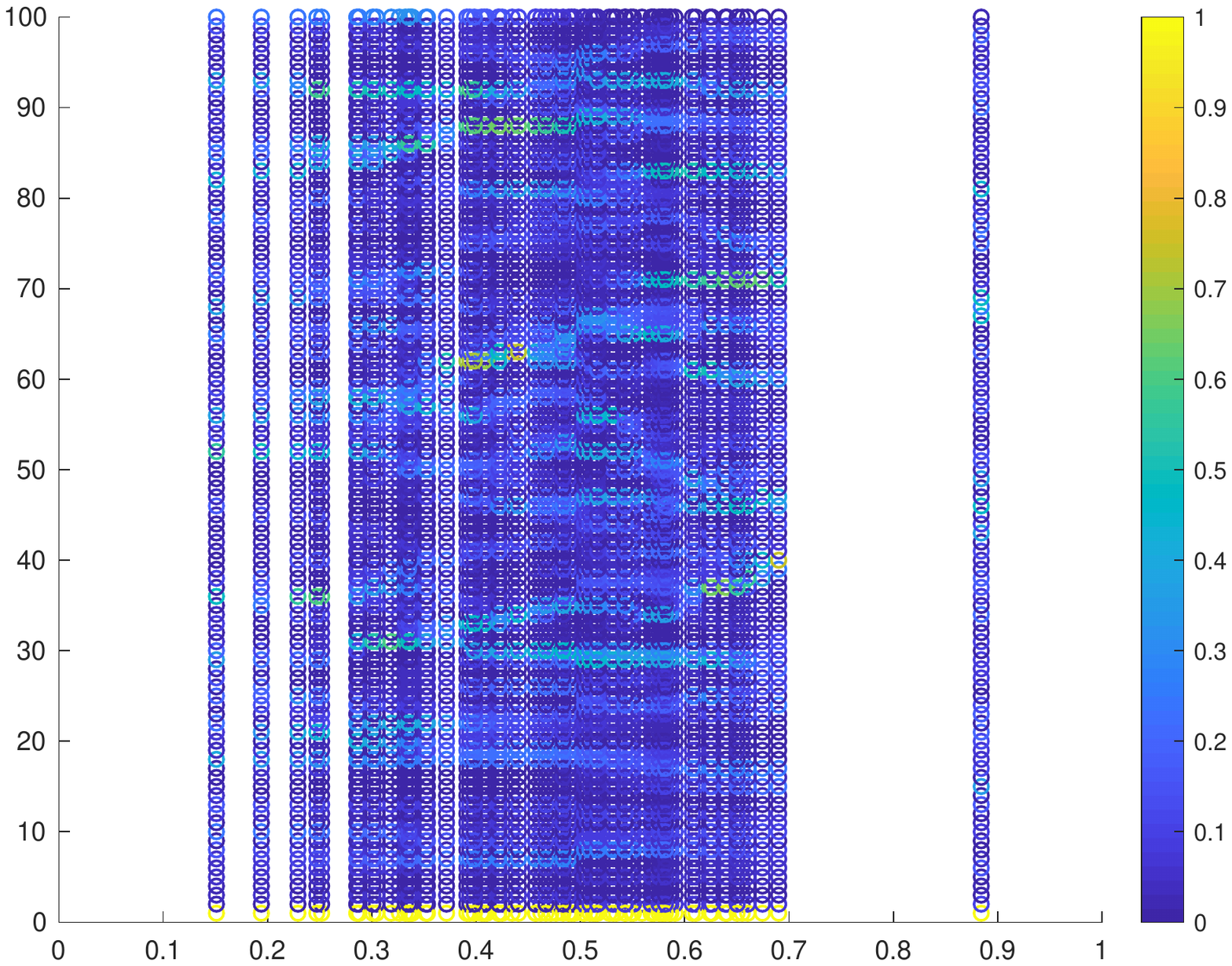}		\caption{Top-left: bandlimited signal, uniform random $x$, top-right: bandlimited signal, Gaussian $x$, bottom-left: random signal, uniform random $x$ and bottom-right: random signal, Gaussian $x$.} \label{fig:het1}
	\end{figure}	
    \end{enumerate}
\end{Example}

From the expression of Fourier transform, $p(x)^{1/2}v_x(i), (x,i)\in X\times [n]$ are the kernels of the integration. However, unlike GSP, they are not pair-wise orthogonal for different $x$.

It is worthwhile to take a closer look at the inverse Fourier transform $\psi_X$. It can be further decomposed as follows:
\begin{align*}
    \psi_X: L^2(X\times [n]) \xrightarrow{\alpha_X} L^2(X\times V) \xrightarrow{\beta_X} L^2(V).
\end{align*}

The function $\alpha_X$ is defined as: the $(x,v)$-component of $\alpha_X(f), f\in L^2(X\times V)$ is the $v$-component of the GSP inverse Fourier transform w.r.t.\ $x$ of $(f(x,i)/p(x)^{1/2})_{i\in [n]}$, where $0/0$ is set to be $0$. The map $\alpha_X$ is clearly invertible a.e.\ by performing fiberwise graph Fourier transforms for $x\in X$ scaled with $p(x)^{1/2}$. The map $\beta_X$ is $\mathbb{E}_{\mu_X}(f(\cdot,v))$. The map $\alpha_X$ is a twisted form of inverse GFT due to our definition of $\phi_X$ and the requirement that $\beta_X\circ \alpha_X = \psi_X$. In subsequent sections, we are going to introduce two filter families:
\begin{enumerate}[(a)]
    \item the \emph{shift invariant family} that ``inserts'' a transformation on $L^2(X\times [n])$ before applying $\alpha_X$; and
    \item a \emph{base change family} that ``inserts'', between $\alpha_X$ and $\beta_X$, a transformation $L^2(X\times V)\to L^2(Y\times V)$, with $Y=X$ or possibly a different probability space.
\end{enumerate}

\section{Convolution filters} \label{sec:conv}

In this section, we shall introduce the family of convolution filters and some important sub-families. 
\begin{Definition}
Given $\Gamma \in L^2(X\times [n])$, multiplication by $\Gamma$ induces $L^2(X\times [n]) \to L^2(X\times [n])$, which is also denoted by $\Gamma$. In turn, we have a \emph{convolution filter} $\Gamma*: L^2(V) \to L^2(V)$ defined by the composition $\psi_X\circ \Gamma\circ\phi_X$. 
\end{Definition}

For each $x\in X$, we use $\Gamma_x$ to denote the projection to $L^2([n])$ via the canonical formula $\Gamma_x(i) = \Gamma(x,i)$. The usual GSP theory gives rises to a fiberwise convolution filter ${\Gamma_x*}$ on $L^2(V)$. It is nothing but the $x$-component of the composition $\alpha_X\circ \Gamma \circ \phi_X$. 

\begin{Example}
	\begin{enumerate}[(a)]
		\item For any signal $g \in L^2(V)$, $\phi_X(g) \in L^2(X\times [n])$. It induces a convolution filter $g*: L^2(V) \to L^2(V)$ that sends $f$ to $\psi_X(\phi_X(g)\cdot\phi_X(f))$. 
		
		\item If there is a uniform upper bound $B$ on the operator norm of $x\in X$, then $\Lambda: (x,i) \mapsto \lambda_x(i)$ belongs to $L^2(X\times [n])$. Moreover, assume that $p(x)$ is the density of $\mu_X$ w.r.t.\ some base measure. We may perform the following computation 
		\begin{align*}
		{\Lambda*}(f) & = \int_x \sum_{i=1}^n \lambda_x(i)p(x)\langle f, v_x(i)\rangle v_x(i) dx \\
		& = \int_x p(x)x(f)dx.
		\end{align*}
		 Consequently, $\Lambda* = \mathbb{E}_{\mu_X}(x)$. More generally, ${\Lambda^n*} = \mathbb{E}_{\mu_X}(x^n)$. As $\phi_X\circ \psi_X$ is not the identity map, ${\Lambda^n*} \neq ({\Lambda*})^n$. This is different from the usual GSP theory.
	\end{enumerate}
\end{Example}

The examples suggest the following observation.

\begin{Lemma} \label{lem:acf}
If $x\in X$ does not have repeated eigenvalues almost everywhere (a.e.), any convolution filter is $\mathbb{E}(r(x))$, where $r$ is a random variable $X\to M_n(\mathbb{R})$ such that $r(x)$ is a degree $n-1$ polynomial in $x$. 
\end{Lemma}

The lemma suggests that if we look at the fiberwise convolution ${\Gamma_x*}$ of ${\Gamma*}$, it is nothing but a polynomial $r(x)= \sum_{0\leq i\leq n-1}a_i(x)x^i$ of degree at most $n-1$ in $x$ a.e. On the other hand, for each fixed degree $0\leq i\leq n-1$, we may look at the coefficient $a_i(x)$ of the $i$-th monomial for each $r(x)$, which gives rise to a function on $x\in X$. This leads us to introduce the following subspace of convolution filters. 

\begin{Definition}
Suppose $X$ is parametrized by $T\subset \mathbb{R}$ via a homeomorphism $t \in T\mapsto x_t\in X$ such that: either (a) $X, T$ are finite or (b) the induced measure on $T$ is absolutely continuous w.r.t.\ the Lebesgue measure. 

A convolution filter ${\Gamma*}$, with $\Gamma \in L^2(X\times n)$, is called \emph{a bi-polynomial filter} on $X$ if for each $0\leq i\leq n-1$, there is a polynomial $a_i(t)$ in $t$ such that ${\Gamma_{x_t}*} = \sum_{0\leq i\leq n-1}a_i(t)x_t^i$.
\end{Definition}

\begin{Theorem}
Suppose $T$ is a finite set of $\mathbb{R}$ or $T = [a,b]$ a interval (with based measure the Lebesgue measure). If $x \in X$ has no repeated eigenvalues and has uniformly bounded operator norm a.e., then every convolution filter is a bi-polynomial filter.
\end{Theorem}

A useful consequence is that: to learn a convolution filter, one may choose suitable finite subset $T'$ of $T$ and find coefficients of each $a_i(t)$ so that the resulting filter agrees with $\Gamma_{x_t}*, t\in T'$. 

\section{Band-pass filters and sampling} \label{sec:ban}

In this section, with the aid of a special family of convolution filters, namely the band-pass filters, we are going to discuss sampling theory.

Let $Y \subset X\times [n]$ be a subset of such that $Y \cap X\times \{i\}, i\in [n]$ is measurable. Recall that the characteristic function $\chi_Y$ on $Y$ is defined as $\chi_Y(y) = 1$ if $y\in Y$ and $0$ otherwise.

\begin{Definition}
The \emph{band-pass filter} $B_Y$ w.r.t.\ $Y$ is defined as the convolution filter associated with $\chi_Y \in L^2(X\times [n])$. For $\epsilon\geq 0$, the set of \emph{$(Y,\epsilon)$-bandlimited signals} consists of signals $f \in L^2(V)$ such that $\norm{B_Y(f)-f}\leq \epsilon$.
\end{Definition}

It is important to take noted that the filter $B_Y$ is not a projection in general. This means that $B_Y$ may not even have non-zero fixed vectors, i.e., for $\epsilon=0$. Therefore, we are not able to define bandlimited signals as the space of fixed vectors of a band-pass filter as in GSP. As a consequence of our definition, the set of $(Y,\epsilon)$-bandlimited signals may not be a vector space. However, if $\mu_X$ is the delta distribution on $x \in X$ and $\epsilon = 0$, then we recover the theory of band-pass filters and bandlimited signals in GSP. 

We have the following basic observation.

\begin{Lemma} \label{lem:tso}
The set of $(Y,\epsilon)$-bandlimited signals is convex. Moreover, if $\epsilon>0$, then $0$ is an interior point.
\end{Lemma}

As a consequence of \cref{lem:tso}, if $\epsilon>0$ and $V'$ is a proper subset of $V$, then the signal values at $V'$ of a $(Y,\epsilon)$-bandlimited signal $f$ do not uniquely determine $f$. Therefore, for signal recovery from sub-samples, we shall aim for approximation instead of exact recovery.  

As $B_Y$ is the expectation of projections, all of its eigenvalues are contained within the closed interval $[0,1]$. Let $0\leq \lambda_1,\ldots, \lambda_n\leq 1$ be the eigenvalues of $B_Y$ ordered increasingly and $v_1,\ldots, v_n$ be the associated eigenvectors. If $f = \sum_{1\leq i\leq n}a_iv_i$ is a $(Y,\epsilon)$-bandlimited signal, then we may estimate, using orthogonality of $v_i, 1\leq i\leq n$, that $\norm{B_Yf-f}^2 = \sum_{1\leq i\leq n}(1-\lambda_i)^2a_i^2 \leq \epsilon^2$. For $1\leq j \leq n$ such that $\lambda_j\neq 1$, we have $\sum_{1\leq i\leq j}a_i^2\leq \epsilon^2/(1-\lambda_j)^2$. Therefore, if $\lambda_j$ is close to $0$ for some $1\leq j\leq n$, then the components of a signal spanned by $v_1, \ldots, v_j$ have a small contribution.

Let $S_j$ be the subspace of signals spanned by $v_{j+1},\ldots, v_n$. A subset $V_j \subset V$ of size $n-j$ is called a \emph{uniqueness set} if the matrix $G_j$, with entries $V_j$-components of $v_i, j+1\leq i\leq n$, is invertible. Denote the operator norm of $G_j^{-1}$ by $\sigma_j$. 

\begin{Proposition}
Suppose observation of a $(Y,\epsilon)$-bandlimited signal $f$ is made at $V_j$, denoted by $f_{V_j}$. Let $f'$ be the linear combination of $v_{j+1},\ldots, v_n$ with coefficients $G_j^{-1}f_{V_j}$. Then
\begin{enumerate}[(a)]
    \item \label{it:nff} $\norm{f'-f} \leq \epsilon(1+\sigma_j)/(1-\lambda_j)$.
    \item $f'$ is $(Y,\epsilon')$-bandlimited with \begin{align*}\epsilon'= \epsilon(1+2\frac{1+\sigma_j}{1-\lambda_j}).\end{align*}
\end{enumerate}
\end{Proposition}

This result shows that the simple sampling procedure using the eigenbasis of $B_Y$ yields reasonable result if both $\lambda_j$ and $\sigma_j$ are small.

\section{Base change} \label{sec:ba}
So far we have been dealing with convolution filters exclusively. On the other hand, if there is a measurable function of base spaces $h:Y\to X$, it induces other types \emph{base change filters}, which we discuss in this section.

First of all, $h$ induces publlback a $h^*: L^2(X\times V) \to L^2(Y\times V)$ as $h^*(f)(y,v) = f(h(y),v)$. Recall that we have the following decomposition of the identity transform: $ I = \psi_X\circ \phi_X = \beta_X\circ \alpha_X \circ \phi_X$. The co-domain of $\alpha_X$ is $L^2(X\times V)$, and this is where $h^*$ should be inserted into. 

On the other hand, as we have seen, $\Gamma \in L^2(X\times [n])$ induces a convolution filter. The map $h$ pulls it back to a function $\Gamma^h \in L^2(Y\times [n])$ defined by $\Gamma^h(y,i) = \Gamma(h(y),i)$. We introduce the following two types of pullback filters w.r.t.\ $h$ and $\Gamma$.  

\begin{Definition}
The filter $F_{\Gamma,h^*}$ associated with $h: Y \to X$ and $\Gamma \in L^2(X\times [n])$ is defined as the composition \begin{align*}\beta_Y \circ h^* \circ \alpha_X\circ\Gamma\circ \phi_X: L^2(V) \to L^2(V).\end{align*}

On the other hand, the filter $F_{\Gamma^h}$ is defined as the composition \begin{align*} \psi_Y\circ \Gamma^h\circ \phi_Y: L^2(V) \to L^2(V). \end{align*}
\end{Definition}

It is worth pointing out that the definition of both filters does not rely on the probability measure of $X$. Therefore, it suffices to require $X$ to be a measure space and $h^*$ is a map between $L^2$ functions. However, $h$ induces a probability measure $h_*(\mu_Y)$ on $X$ by $h_*(\mu_Y)(X')=\mu_Y(h^{-1}(X'))$ for measurable subset $X'\subset X$, called the pushforward of measure. Therefore, we do not lose generality by assuming both $X$ and $Y$ are probability spaces.

Suppose $\mu_Y$ is absolutely continuous w.r.t.\ the base measure on $Y$ so that it can be written as $qdy$ for density function $q$. Then it is convenient to write down explicitly $F_{\Gamma,h^*}$ and $F_{\Gamma^h}$, for $f\in L^2(V)$, as follows:
\begin{align*}
    F_{\Gamma,h^*}(f) = \int_Y q(y)\sum_{1\leq i\leq n}\Gamma(h(y),i)\langle f, v_{h(y)}(i)\rangle v_{h(y)}(i) dy;
\end{align*}
\begin{align*}
    F_{\Gamma^h}(f) = \int_Y q(y)\sum_{1\leq i\leq n}\Gamma(h(y),i)\langle f, v_{y}(i)\rangle v_{y}(i) dy.
\end{align*}

From these formulas, we have the intuition that $F_{\Gamma,h^*}$ performs a fiberwise convolution and aggregate according to $\mu_Y$. It can be viewed as a ``re-arragements'' of ``probability'' density. However, $F_{\Gamma^h}$ ``re-arranges" the convolution kernel $\Gamma$.

\begin{Example}
\begin{enumerate}[(a)]
    \item Suppose $Y$ is a finite discrete subset of $X$ and $h: Y \to X$ is the inclusion. Then the induced measure on $X$ is a discrete measure supported on $Y$. The pullback of any function, via $h$, is just the restriction to $Y$. In this case, $F_{\Gamma,h^*} = F_{\Gamma^h}$ performs the following: apply a pointwise convolution, with the convolution kernel $\Gamma$ restricted to $y$, at each $y\in Y$; and then take the expectation according to the discrete measure on $Y$. The resulting filter is the same as a sampling procedure.
    \item For convenience, we work with the parameter spaces. Suppose $Y = [0,1]$ has the usual Lebesgue measure and $X = \{x_1,\ldots, x_m\}$ is a finite subset of $Y$ of size $m$. If $Y = \cup_{1\leq i\leq m} Y_i$ has a decomposition into disjoint intervals such that $x_i \in Y_i$. The map $h: Y \to X$ sends the interval $Y_i$ to $x_i, 1\leq i\leq m$. The induced measure on $X$ assigns length of $Y_i$ to $x_i$. If $\Gamma = (\Gamma_i \in \mathbb{R}^n)_{1\leq i\leq m}$ is a function on $X\times [n]$, then the filter $F_{\Gamma,h^*}$ performs the following: apply a pointwise convolution, with convolution kernel $\Gamma_i$, at each $x_i$; and then take the expectation as the weighted sum with weights length of $Y_i,1\leq i\leq m$. This is the \emph{coarsening procedure}. In general, it is not a convolution filter on $Y$ and hence $F_{\Gamma,h^*}\neq F_{\Gamma^h}$. 
    \item Suppose $X = Y = [0,1]$ have the Lebesgue measure. Let $G$ be a square lattice and $L_1, L_2$ are the Laplacians of the subgraphs consisting of vertical and horizontal edges respectively. As earlier, for $x\in [0,1]$, we have the matrix $L_x = xL_1+(1-x)L_2$. For $0<\eta$, define $h_{\eta}: Y \to X$ by the formula $h_{\eta}(y) = y\eta/(1-y+y\eta) \in [0,1]$. The map $h_{\eta}$ is invertible with inverse given by $x \mapsto x/(x+\eta-x\eta)$. It can be verified that if $H_x= xL_1 + (1-x)(\eta L_2)$ as a stretched version of $L_x$, then $H_x$ is a scaler multiple of $L_y$ for $x = h_{\eta}(y)$. As a consequence, suppose our knowledge on the graph distribution is on the unstretched version $L_y$ and the signal is stretched in the horizontal direction by a factor $\eta$. Then to match prior knowledge and observed signal in a convolution process, one needs to use the filter with base change $F_{\Gamma,h^*}$. An illustration is shown in \figref{fig:bc}
    	\begin{figure}[!htb] 
		\centering
		\includegraphics[width=0.45\columnwidth]{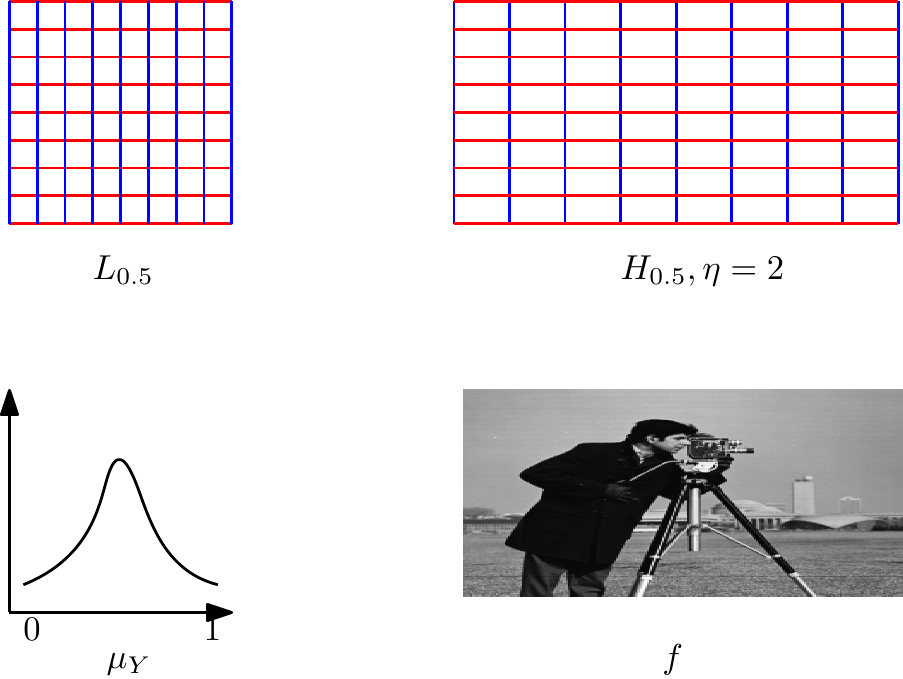}
		\caption{The figures show the case $\eta=2$, i.e., the graph is stretched horizontally by a factor of $2$. Suppose the signal $f$, e.g.\ the image show below, is stretched and knowledge of the graph distribution is based on unstretched graphs. To correctly perform signal processing, one needs to apply base change.} \label{fig:bc}
	\end{figure}	
\end{enumerate}
\end{Example}

\section{Simulation results} \label{sec:sim}
The framework discussed in this paper relies on knowing the probability space $(X,\mu_X)$ on $V$. When such information is not directly available, we propose a Bayesian learning framework to learn $(X,\mu_X)$ with a small amount of training data. For this, we follow largely the treatment in \cite{Gue19}.

Suppose we have a candidate space $X$, and want to learn a discrete distribution $\mu_X$ that approximates the true distribution. We need the following data: 
\begin{itemize}
\item There is a set of training signals: $D_k = \{f_i \in L^2(V), 1\leq i\leq k\}$, possible with labels $Z_k = \{z_i \in \mathbb{R}, 1\leq i\leq k\}$. 
\item There is a loss function required to be minimized: $\ell: M_n(\mathbb{R})\times L^2(V) \to \mathbb{R}_+$. In the labelled case, $\ell$ should have domain $M_n(\mathbb{R})\times L^2(V)\times \mathbb{R}$.

\item There is a prior distribution $\mu_0$, e.g., the uniform distribution.
\end{itemize}

For $x\in X$, we may now define the empirical risk as
\begin{align*}
    r(x) = \frac{1}{k} \sum_{1\leq i\leq k}\ell(x,f_i), \text{ or } r(x) = \frac{1}{k} \sum_{1\leq i\leq k}\ell(x,f_i,z_i).
\end{align*}

For a fixed parameter $\gamma > 0$, discrete samples to approximate $\mu_X$ are drawn proportional to $\exp\big(-\gamma r(\cdot) \big)\mu_0$, yielding the Gibbs posterior. One of the most important methods to generate such samples is the Metropolis-Hastings algorithm.   

The network is a weather station network in the United States with $|V|=197$ nodes.\footnote{http://www.ncdc.noaa.gov/data-access/land-based-station-data/station-metadata} The signals considered  are daily temperatures recorded over the year 2013.\footnote{ftp://ftp.ncdc.noaa.gov/pub/data/gsod} We consider two signal processing problems: \emph{sampling} and \emph{anomaly detection}.

\begin{figure}[!htb] 
		\centering
		\includegraphics[width=0.45\columnwidth,trim=0 6cm 0 6cm, clip]{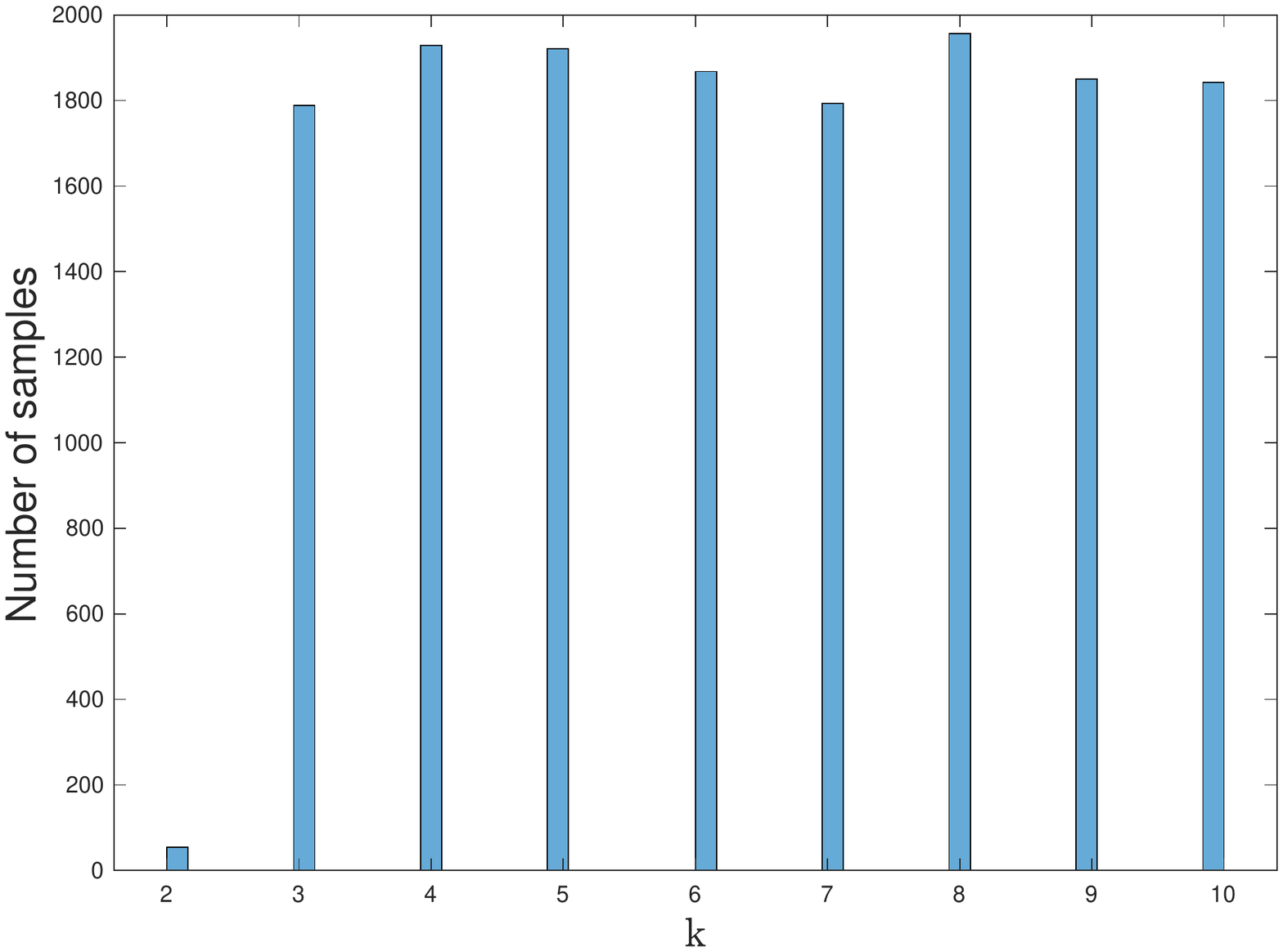}
		\includegraphics[width=0.45\columnwidth,trim=0 6cm 0 6cm, clip]{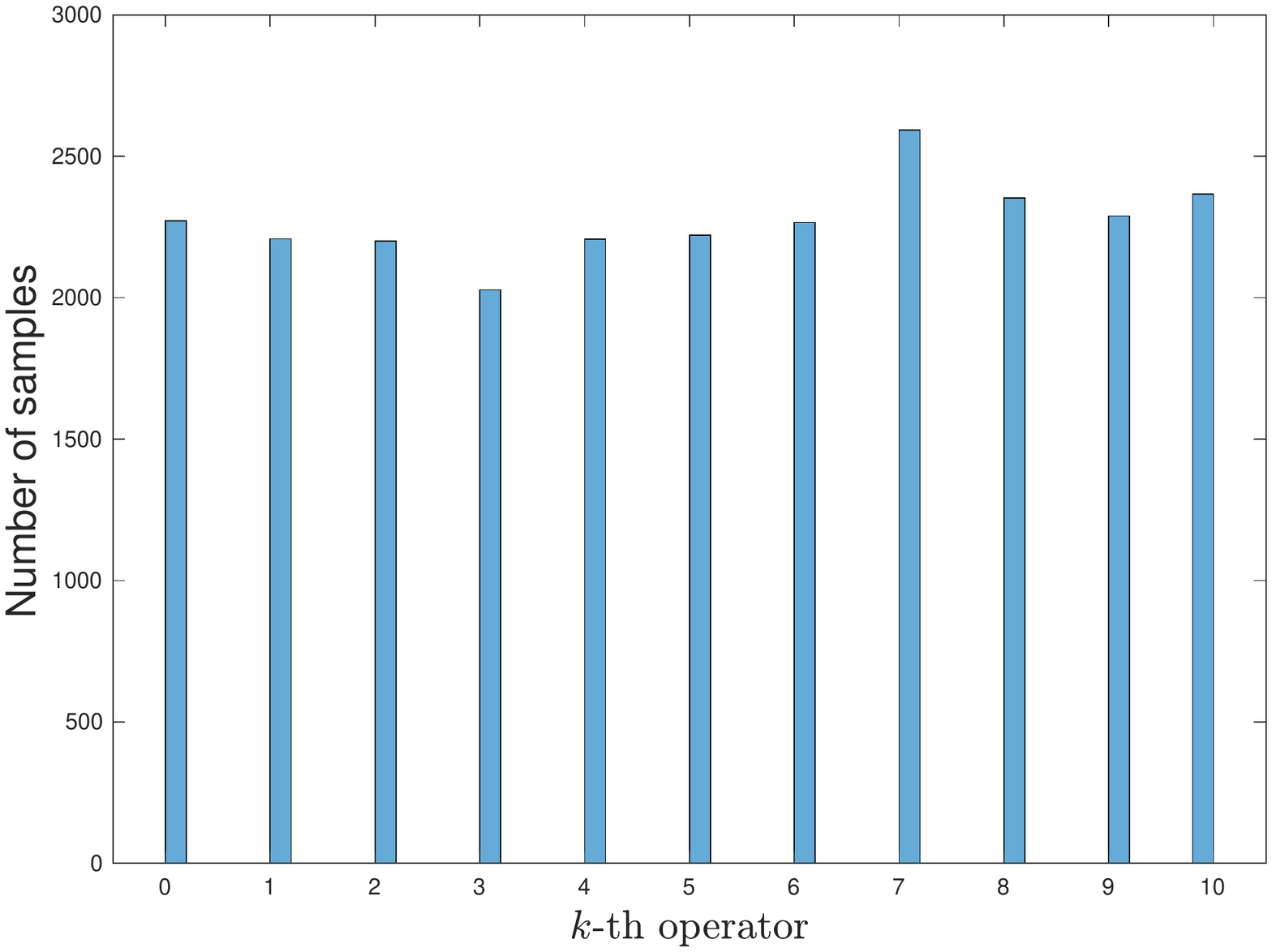}
		\caption{Empirical distributions: left for sampling, and right for anomaly detection.} \label{fig:dsp1}
	\end{figure}

For sampling, we aim to sample $10$ stations and use the reading at these stations to recover the reading over the entire network as described in \cref{sec:ban}. The space $X$ are parameterized by $k=2,\ldots, 10$. For each $k$, we construct the $k$-NN graph $G_k$ and obtain the Laplacian $x_k=L_k$. To learn a distribution on $X$, let $\hat{f}_k$ be the usual GFT of a signal w.r.t.\ $x_k$. We define $\ell(x_k,f)$ as the $\sqrt{\sum_{11\leq j\leq 197}\hat{f}_k(j)^2}/\norm{f}$, as the ``energy of high frequency components''. The training use $30$ signals. The resulting empirical distribution is shown in \figref{fig:dsp1}. We see that $k=2$ has small weight, and hence run signal recovery from sample reading for $k=3,\ldots, 10$. The average recovery error $\epsilon$ is shown in \cref{tab:dsp1}, where $Y= X\times [10]$ in $B_Y$. We see that working with a distribution yields the best result. Moreover, it is hard to determine from the empirical distribution the single best graph to choose.

\begin{table}[!htb]
\centering 
\caption{Sampling error} \label{tab:dsp1}
\scalebox{0.88}{\begin{tabular}{|l|c|c|c|c|c|c|c|c|c|} 
\hline
 $k$ & $3$ & $4$ & $5$ & $6$ & $7$ & $8$ & $9$ & $10$ & $B_Y$\\ 
\hline\hline
\emph{$\epsilon$} & $1.55$ & $3.07$ & $267$ & $318$ & $18.8$ & $1.54$ & $1.56$ & $169$ & $1.46$\\
\hline
\end{tabular}}
\end{table}   

For the anomaly detection, we follow the setup of \cite{JGT20}. In particular, $X$ consists of $11$ operators based on $11$ simplicial structures. The loss function $\ell(x,f)=1$ if $x$ successfully tells $f$ is abnormal as in \cite{JGT20} Section VI B, and $0$ otherwise. The empirical distribution is shown in \figref{fig:dsp1}. We perform anomaly detection for reading perturbation chosen from $40-60$ degrees. The successful detection rate $R$ is show in \cref{tab:dsp2}, with the last entry from our framework. Again, we see that working with a distribution directly results the best performance, and a single best graph is hard to determine.  

\begin{table}[!htb]
\centering 
\caption{Successful anomaly detection rate} \label{tab:dsp2}
\scalebox{0.65}{\begin{tabular}{|l|c|c|c|c|c|c|c|c|c|c|c|c|} 
\hline
 Index & $0$ & $1$ & $2$ & $3$ & $4$ & $5$ & $6$ & $7$ & $8$ & $9$ & $10$ & \\ 
\hline\hline
\emph{$R$ ($\%$)} & $74.3$ & $66.1$ & $53.6$ & $63.3$ & $74.2$ & $75.3$ & $67.3$ & $68.0$ & $66.4$ & $67.6$ & $68.9$ & $76.9$ \\
\hline
\end{tabular}}
\end{table}   

\section{Conclusions} \label{sec:con}

In this paper, we present a new signal processing framework with a distribution of graph operators. We define Fourier transform and discuss the theory of filtering and sampling. The usefulness of the framework is demonstrated with simulations on real dataset. For future work, we shall explore the possibility of using the framework to develop new graph neural network (GNN) methods.   

\newpage


\bibliographystyle{IEEEbib}
\bibliography{IEEEabrv,strings,allref}

\end{document}

%% file: preamble.tex
\usepackage[T1]{fontenc}
\usepackage{amsmath,amssymb,amsfonts,mathrsfs,bm}
\usepackage{mathtools}
\usepackage{amsthm}
\usepackage{nicefrac}
\usepackage{cite}
\usepackage[shortlabels]{enumitem}
\usepackage{graphicx}
\usepackage{epstopdf}
\usepackage{url}
\usepackage{colortbl}
\usepackage{booktabs}
\usepackage{multirow}
\usepackage[table,dvipsnames]{xcolor}
\usepackage[normalem]{ulem}

\usepackage{array}
\newcolumntype{L}[1]{>{\raggedright\let\newline\\\arraybackslash\hspace{0pt}}m{#1}}
\newcolumntype{C}[1]{>{\centering\let\newline\\\arraybackslash\hspace{0pt}}m{#1}}
\newcolumntype{R}[1]{>{\raggedleft\let\newline\\\arraybackslash\hspace{0pt}}m{#1}}

\makeatletter
\let\MYcaption\@makecaption
\makeatother
\usepackage[font=footnotesize]{subcaption}
\makeatletter
\let\@makecaption\MYcaption
\makeatother

\usepackage{xparse}



\usepackage{glossaries}

\makeatletter
\let\oldgls\gls
\let\oldglspl\glspl

\newcommand\fussy@ifnextchar[3]{%
  \let\reserved@d=#1%
  \def\reserved@a{#2}%
  \def\reserved@b{#3}%
  \futurelet\@let@token\fussy@ifnch}
\def\fussy@ifnch{%
  \ifx\@let@token\reserved@d
    \let\reserved@c\reserved@a 
  \else
    \let\reserved@c\reserved@b
  \fi
 \reserved@c}

\renewcommand{\gls}[1]{%
  \oldgls{#1}\fussy@ifnextchar.{\@checkperiod}{\@}}
\renewcommand{\glspl}[1]{%
  \oldglspl{#1}\fussy@ifnextchar.{\@checkperiod}{\@}}

\newcommand{\@checkperiod}[1]{%
  \ifnum\sfcode`\.=\spacefactor\else#1\fi
}
\makeatother

\newacronym{wrt}{w.r.t.}{with respect to}
\newacronym{RHS}{RHS}{right-hand side}
\newacronym{LHS}{LHS}{left-hand side}
\newacronym{iid}{i.i.d.}{independent and identically distributed}

\usepackage{float}

\ifx\notloadhyperref\undefined
	\ifx\loadbibentry\undefined
		\usepackage[hidelinks,hypertexnames=false]{hyperref} 
	\else
		\usepackage{bibentry}
		\makeatletter\let\saved@bibitem\@bibitem\makeatother
		\usepackage[hidelinks,hypertexnames=false]{hyperref}
		\makeatletter\let\@bibitem\saved@bibitem\makeatother
	\fi
\else
	\relax
\fi

\usepackage[capitalize]{cleveref}
\crefname{equation}{}{}
\Crefname{equation}{}{}
\crefname{claim}{claim}{claims}
\crefname{step}{step}{steps}
\crefname{line}{line}{lines}
\crefname{condition}{condition}{conditions}
\crefname{dmath}{}{}
\crefname{dseries}{}{}
\crefname{dgroup}{}{}

\crefname{Problem}{Problem}{Problems}
\crefformat{Problem}{Problem~(#2#1#3)}
\crefrangeformat{Problem}{Problems~(#3#1#4) to~(#5#2#6)}

\crefname{Theorem}{Theorem}{Theorems}
\crefname{Corollary}{Corollary}{Corollaries}
\crefname{Proposition}{Proposition}{Propositions}
\crefname{Lemma}{Lemma}{Lemmas}
\crefname{Definition}{Definition}{Definitions}
\crefname{Example}{Example}{Examples}
\crefname{Assumption}{Assumption}{Assumptions}
\crefname{Remark}{Remark}{Remarks}
\crefname{Rem}{Remark}{Remarks}
\crefname{remarks}{Remarks}{Remarks}
\crefname{Appendix}{Appendix}{Appendices}
\crefname{Exercise}{Exercise}{Exercises}
\crefname{Theorem_A}{Theorem}{Theorems}
\crefname{Corollary_A}{Corollary}{Corollaries}
\crefname{Proposition_A}{Proposition}{Propositions}
\crefname{Lemma_A}{Lemma}{Lemmas}
\crefname{Definition_A}{Definition}{Definitions}

\usepackage{crossreftools}
\usepackage{algorithm,algorithmic}

\ifx\loadbreqn\undefined
	\relax
\else
	\usepackage{breqn} 
\fi

\ifx\renewtheorem\undefined
\ifx\useTheoremCounter\undefined
\newtheorem{Theorem}{Theorem}
\newtheorem{Corollary}{Corollary}
\newtheorem{Proposition}{Proposition}
\newtheorem{Lemma}{Lemma}
\else
\newtheorem{Theorem}{Theorem}

\newtheorem{Proposition}[theorem]{Proposition}
\fi

\newtheorem{Definition}{Definition}
\newtheorem{Example}{Example}

\fi

\theoremstyle{remark}

\theoremstyle{plain}

\DeclareSymbolFont{bsfletters}{OT1}{cmss}{bx}{n}
\DeclareSymbolFont{ssfletters}{OT1}{cmss}{m}{n}
\DeclareMathSymbol{\bsfGamma}{0}{bsfletters}{'000}
\DeclareMathSymbol{\ssfGamma}{0}{ssfletters}{'000}
\DeclareMathSymbol{\bsfDelta}{0}{bsfletters}{'001}
\DeclareMathSymbol{\ssfDelta}{0}{ssfletters}{'001}
\DeclareMathSymbol{\bsfTheta}{0}{bsfletters}{'002}
\DeclareMathSymbol{\ssfTheta}{0}{ssfletters}{'002}
\DeclareMathSymbol{\bsfLambda}{0}{bsfletters}{'003}
\DeclareMathSymbol{\ssfLambda}{0}{ssfletters}{'003}
\DeclareMathSymbol{\bsfXi}{0}{bsfletters}{'004}
\DeclareMathSymbol{\ssfXi}{0}{ssfletters}{'004}
\DeclareMathSymbol{\bsfPi}{0}{bsfletters}{'005}
\DeclareMathSymbol{\ssfPi}{0}{ssfletters}{'005}
\DeclareMathSymbol{\bsfSigma}{0}{bsfletters}{'006}
\DeclareMathSymbol{\ssfSigma}{0}{ssfletters}{'006}
\DeclareMathSymbol{\bsfUpsilon}{0}{bsfletters}{'007}
\DeclareMathSymbol{\ssfUpsilon}{0}{ssfletters}{'007}
\DeclareMathSymbol{\bsfPhi}{0}{bsfletters}{'010}
\DeclareMathSymbol{\ssfPhi}{0}{ssfletters}{'010}
\DeclareMathSymbol{\bsfPsi}{0}{bsfletters}{'011}
\DeclareMathSymbol{\ssfPsi}{0}{ssfletters}{'011}
\DeclareMathSymbol{\bsfOmega}{0}{bsfletters}{'012}
\DeclareMathSymbol{\ssfOmega}{0}{ssfletters}{'012}

\newcommand{\qednew}{\nobreak \ifvmode \relax \else
      \ifdim\lastskip<1.5em \hskip-\lastskip
      \hskip1.5em plus0em minus0.5em \fi \nobreak
      \vrule height0.75em width0.5em depth0.25em\fi}

\newcommand{\norm}[1]{{\left\lVert{#1}\right\rVert}}

\DeclareDocumentCommand\set{s m t| m}{%
  \IfBooleanTF#1%
	{\left\{\, #2\mathrel{} \IfBooleanTF{#3}{\middle|}{:}\mathrel{}  #4\, \right\}}%
  {\{\, #2 \IfBooleanTF{#3}{\mid}{\mathrel{} : \mathrel{}} #4\, \}}%
}

\DeclareDocumentCommand \ifcond {m m} {%
	{#1} %
	\IfValueT{#2}{\, \middle|\, {#2}}%
}

\DeclareDocumentCommand \P {e{_} g >{\SplitArgument{ 1 }{ @| }}d() g } {%
	\mathbb{P}%
	\IfValueTF{#1}{_{#1}}
		{\IfValueT{#2}{_{#2}}}%
	\IfValueT{#3}{\left(\ifcond#3}%
	\IfValueT{#4}{\, \middle|\, {#4}}%
	\IfValueT{#3}{\right)}%
}

\DeclareDocumentCommand \E {e{_} g >{\SplitArgument{ 1 }{ @| }}o g } {%
	\mathbb{E}%
	\IfValueTF{#1}{_{#1}}
		{\IfValueT{#2}{_{#2}}}%
	\IfValueT{#3}{\left[\ifcond#3}%
	\IfValueT{#4}{\, \middle|\, {#4}}%
	\IfValueT{#3}{\right]}%
}

\definecolor{gray90}{gray}{0.9}

\ifx\nohighlights\undefined

	\newcommand{\msout}[1]{\text{\color{green} \sout{\ensuremath{#1}}}}
	\newcommand{\del}[1]{{\color{green}\ifmmode \msout{#1}\else\sout{#1}\fi}}
\else

	\newcommand{\msout}[1]{#1}
	\newcommand{\del}[1]{#1}
\fi

\newcommand{\hhide}[1]{}

\renewcommand{\figurename}{Fig.}
\newcommand{\figref}[1]{\figurename~\ref{#1}}
\graphicspath{{./Figures/}} 
\ifx\diagnoselabel\undefined
	\relax
\else
	\makeatletter
	 \def\@testdef #1#2#3{%
		 \def\reserved@a{#3}\expandafter \ifx \csname #1@#2\endcsname
		\reserved@a  \else
	 \typeout{^^Jlabel #2 changed:^^J%
	 \meaning\reserved@a^^J%
	 \expandafter\meaning\csname #1@#2\endcsname^^J}%
	 \@tempswatrue \fi}
	\makeatother
\fi
